\documentclass[twocolumn,prb,showpacs]{revtex4}
\usepackage{amssymb}
\usepackage{amsmath}
\usepackage{graphicx}
\usepackage{dcolumn}
\usepackage{verbatim}
\usepackage{bm}
\usepackage{color}
\usepackage{epstopdf}

\begin{document}

\title{Interplay between the Fulde-Ferrell phase and Larkin-Ovchinnikov phase in the superconducting ring pierced by an Aharonov-Bohm flux}

\author{H. T. Quan and Jian-Xin Zhu}
\affiliation{Theoretical Division, Los Alamos National Laboratory, Los Alamos,
New Mexico 87545, U.S.A.}
\date{\today}

\begin{abstract}
We study the phase diagram of a superconducting ring threaded by an Aharonov-Bohm flux and an in-plane magnetic Zeeman field. The simultaneous presence of both the external flux and the in-plane magnetic field leads to the competition between the Fulde-Ferrell (FF) phase and the Larkin-Ovchinnikov (LO) phase. Using the Bogoliubov-de Gennes equation, we investigate the spacial profile of the order parameter. Both the FF phase and the LO phase are found to exist stably in this system. The phase boundary is determined by comparing the free energy. The distortion of the phase diagrams due to the mesoscopic effect is also studied.
\end{abstract}

\maketitle

\section{introduction}

In recent years, one of the inhomogeneous superconducting states, known as Fulde-Ferrell-Larkin-Ovchinnikov (FFLO) state, has received a lot of interest. This superconducting state with periodical spacial variation of order parameter (OP) was first proposed independently by Fulde and Ferrell~\cite{FF64} and by Larkin and Ovchinnikov~\cite{LO64} in 1960s. The possible evidence of its existence has been reported in certain unconventional
superconductors~\cite{HARadovan03,ABianchi03} and the possibility of its realization in trapped cold atoms.~\cite{MWZwierlein06,GBPartridge06,XJLiu07,RSharma08,WLLu09} In literature, the state is collectively known as the FFLO state.~\cite{YMatsuda07} Actually, they are two kinds of states with slight difference: the order parameter of the LO  state is real and spatially inhomogeneous, which breaks the translational symmetry, while the order parameter of the FF state has a uniform magnitude, but an inhomogeneous phase similar to that of a plane wave, breaks the time-reversal symmetry. According to previous studies, the FF state is usually unstable, and unfavorable in comparison with the LO state.
Although the LO to FF phase transition was predicted in Ref.~\onlinecite{KYang01}, a more recent study~\cite{QWang07} shows that there is no stable FF phase in such a system and there is no LO to FF phase transition either. The authors in Ref. \onlinecite{QCui06}, mention a possible FF state in a momentum space study, but as to the best of our knowledge, a realization of stable FF state in the presence of a Zeeman field has not been reported yet in a real space calculation.

As is well known when a Zeeman field is added to a superconductor, the LO state becomes favorable in comparison with the BCS state, irrespective of the geometry of the superconductor. Meanwhile, we notice that in a superconducting ring, which is threaded by a magnetic flux, the Aharonov-Bohm (AB) flux breaks the time reversal symmetry in much the same spirit as that in the FF phase.~\cite{JXZhu94} As a result the FF state comes out. An interesting question is then if we add both the magnetic flux and an in-plane magnetic field, how will the two phases compete with each other? Motivated by this observation, we study in this paper the interplay between this AB flux-driven FF phase and the Zeeman field-induced LO phase. It is of great interest to study the phase transitions and phase diagram in such a system. The investigation is carried out in a tight-binding model for a superconducting ring pierced by an AB magnetic flux, and in the presence of a Zeeman magnetic field. We solve self-consistently the Bogliubov de Gennes equation for the superconducting order parameter and determine the phase diagram by comparing the total energy. We find that for this system, there are four different phases when we vary the two parameters, magnetic flux $\Phi$ and the Zeeman field $h$. More interestingly, we also study the mescscopic effect.

The paper is organized as follows: in Sec.~\ref{sec:method}, we introduce the tight-binding model and present the mean-field treatment. In Sec.~\ref{sec:result}, we numerically carry out the calculation of superconducting order parameter as a function of the magnetic flux and Zeeman field, and determine the phase diagram by comparing the free energies. Section~\ref{sec:discussion} is the discussion and conclusion.

\section{Model and mean-field treatment}
\label{sec:method}

\begin{figure}[ht]
\begin{center}
\includegraphics[bb=110 326 471 619, width=6.0 cm, clip]{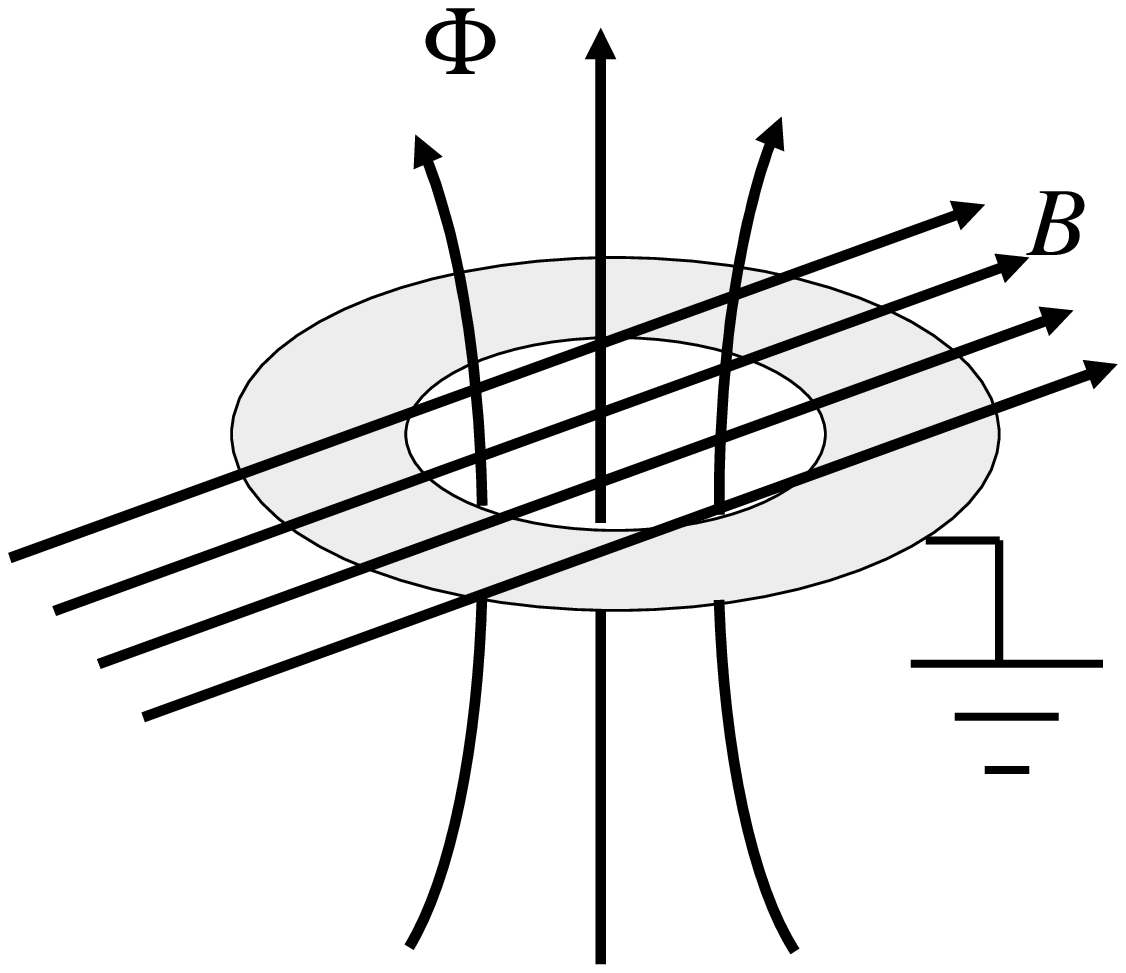}
\end{center}
\caption{(Color online) Schematic illustration of the setup. A  superconducting ring is threaded by an external magnetic flux, denoted by $\Phi$. A magnetic field $B$ is applied in the plane of the ring. The ring is connected to the ground to ensure that the chemical potential is fix, but the electron number may fluctuate}
\label{fig1}
\end{figure}

We consider a one-dimensional superconducting ring threaded by an external magnetic flux $\Phi$ (see Fig. \ref{fig1}). Meanwhile, there is an in-plane magnetic field $B$, which generates the Zeeman spliting and gives rise to the inhomogeneous pairing. The system is described by the following Hamiltonian
\begin{equation}
\begin{split}
H=& - \sum_{i, j, \sigma} \tilde{t}_{ij} c_{i \sigma}^{\dagger} c_{j \sigma} +h \sum_{i, \sigma} \sigma c_{i \sigma}^{\dagger} c_{i \sigma} \\
&-V \sum_{i} n_{i \uparrow}n_{i \downarrow} - \mu \sum_{i, \sigma} c_{i \sigma}^{\dagger} c_{i \sigma}. \label{1}
\end{split}
\end{equation}
Here $\tilde{t}_{ij}=t_{ij} e^{i 2\pi \Phi/N \Phi_{0}}$, where $t_{ij}$ is the bare hopping coefficient and $\Phi_{0}=hc/e$ is the normal-state flux quantum, and $N$ is number of lattice sites for the ring; $c_{i \sigma}^{\dagger}$ ($c_{i \sigma}$) is the creation (annihilation) operator on the $i$-th lattice site with spin $\sigma=\pm 1$ for spin up and down electrons, arising from the interaction between the magnetic field and the spin of the electrons; $n_{i,\sigma}=c_{i \sigma}^{\dagger} c_{i \sigma}$ is the particle number on the $i$-th site with spin $\sigma$; $g$ is equal to 2; $\mu_{B}$ is the Bohr magneton and $B$ is the strength of the in-plane magnetic field; $V$ is the strength of the on-site pairing interaction; $\mu$ is the chemical potential. For simplicity, we define $h=g\mu_{B} B$ as the strength of the Zeeman field. In the present work, we take $t_{ij}$ to be $t$ between nearest neighboring sites and zero otherwise. Within the mean-field approximation, the Hamiltonian (\ref{1}) is reduced to
\begin{equation}
\begin{split}
H=& - \sum_{i, j, \sigma} \tilde{t}_{ij} c_{i \sigma}^{\dagger} c_{j \sigma} +h \sum_{i, \sigma} \sigma c_{i \sigma}^{\dagger} c_{i \sigma} - \mu \sum_{i, \sigma} c_{i \sigma}^{\dagger} c_{i \sigma} \\
&+V \sum_{i} (\Delta_{i} c_{i \uparrow}^{\dagger} c_{i \downarrow}^{\dagger} +h.c.) +\sum_{i} \frac{\left\vert \Delta_{i} \right\vert ^2}{V},  \label{2}
\end{split}
\end{equation}
where $\Delta_{i} \equiv V \left\langle c_{i \uparrow} c_{i \downarrow} \right\rangle$ is the pair potential. To diagonalize this Hamiltonian, we employ the following Bogoliubov transformation
\begin{equation}
\begin{split}
c_{i \sigma}=&\sum_{\nu} \left[ u_{i \sigma}^{\nu} \gamma_{\nu} - \sigma (v_{i \sigma}^{\nu}) ^{\ast} \gamma_{\nu} ^{\dagger} \right] \\
c_{i \sigma}^{\dagger}=&\sum_{\nu}  \left[ (u_{i \sigma}^{\nu}) ^{\ast} \gamma_{\nu} ^{\dagger} - \sigma v_{i \sigma}^{\nu} \gamma_{\nu}  \right], \label{3}
\end{split}
\end{equation}
corresponding to the eigenvalues $E_{\nu}^{\sigma}$ where $\gamma_{\nu}$ and $\gamma_{\nu}^{\dagger}$ are the quasi-particle operators. The coefficients ($u_{i \sigma}^{\nu}, v_{i \sigma}^{\nu}$) satisfy the Bogoliubov-de Gennes (BdG) equation:~\cite{PGdeGennes65}
\begin{equation}
\sum_{j} \left[
\begin{array}{cc}
 H_{ij\sigma} &   \Delta_{i} \delta_{ij} \\
 (\Delta_{i})^{\ast} \delta_{ij}, & - \overline{H}_{ij\sigma} \\
\end{array}
\right]
\left[
\begin{array}{c}
u_{j \sigma}^{\nu}  \\
v_{j \overline{\sigma}}^{\nu} \\
\end{array}
\right]
= E_{\nu}^{\sigma}
\left[
\begin{array}{c}
u_{i \sigma}^{\nu}  \\
v_{i \overline{\sigma}}^{\nu} \label{4}
\end{array}
\right]
\end{equation}
where $H_{ij\sigma}=- \tilde{t}_{ij} - \mu \delta_{ij}+\sigma h \delta_{ij} $, and $\overline{H}_{ij\sigma}=[- \tilde{t}_{ij} - \mu \delta_{ij} ]^{\ast} +  \overline{ \sigma} h^{\ast}  \delta_{ij} $. The self consistent equation of the pair potential
\begin{equation}
\Delta_{i}=\frac{V}{2} \sum_{\nu=1}^{2N} u_{i \uparrow}^{\nu} (v_{i \downarrow}^{\nu}) ^{\ast} \tanh {\frac{E_{\nu}^{\uparrow}}{2T}}  \label{5}
\end{equation}
is solved by iteration. Here  $T$ is the temperature the Boltzmann constant $k_{B}=1$ has been taken.) . Notice that the quasiparticle energy is measured with respect to the chemical potential.

\section{Numerical results}
\label{sec:result}

In our numerical calculation, we take the energy unit $t=1$, and the chemical potential $\mu=-0.5$, the interaction strength $V=2$, and the ring size $N=50$. Though the system size is far from the thermodynamic limit, it already gives the same phase boundary as  that of infinite $N$. The order parameter structure depends not only on the Zeeman field $h$, but also the magnetic flux $\Phi$. 
We note~\cite{JXZhu94,NByers62} that all physical quantities have already been a function of $\Phi$ with a period of $\Phi_0$ even in the normal state.
Therefore, it is sufficient for us to consider the magnetic flux in the range $\Phi \in [0,\Phi_{0}]$. In the absence of the magnetic flux $\Phi=0$, the Bardeen-Cooper-Schrieffer (BCS) order parameter $\Delta=0.351$ for $h=0$, and the LO state is stable for $h_{c1}<h<h_{c2}$ with $h_{c1}=0.23$ and $h_{c2}=1.56$. The system becomes normal ($\Delta=0$) for $h>h_{c2}$. In the presence of the magnetic flux, the magnetic flux can induce a change in the structure of the BCS state in an $s$-wave superconductor, namely a crossover from the BCS state in the absence of a magnetic flux to a FF state with a magnetic flux when the Zeeman field is low $h<h_{c1}$. When the Zeeman field increases, the LO becomes favorable and both BCS and FF states give in. If we continue to increase the Zeeman field, the amplitude of the pairing potential of the LO phase will be suppressed  by the Zeeman field until it disappears finally, and the system enters the normal state. In the following, we will numerically construct the phase diagrams.

\subsection {Phase boundary in $\emph{h}$-$\Phi$ plane}

We first focus on the low temperature case $\beta=1/ T=200$ (corresponding to $T=0.005$). In the absence of the magnetic flux, there are three different phases: BCS, LO, and normal. In the presence of the magnetic flux, there are also three phases, FF, LO, and normal state. In the following, we study the phase transitions and the phase boundaries for fixed temperature when varying $h$ and $\Phi$.

In order to check if the FF state becomes the ground state, we assign a periodic phase to the order parameter at each site as an initial condition. Similarly, we assign a constant phase to see if BCS state becomes the ground state. For a set of fixed parameters ($h$, $\Phi$, $N$, $T$), different stable  solutions (with different order parameter textures) could be obtained from different initial configurations. For example, one may find both stable LO-type OP and FF-type OP for the same set of parameters ($N=50,\; \beta=200,\; \Phi=0.25 \Phi_{0},\; h=0.25$). Even there are more than one stable LO type solutions for the same set of parameters, which means different net momentum of the Cooper pair. To distinguish one state from other competing states (including BCS state and FF state), we choose the energetically most favored one by comparing their free energies. For the model (\ref{1}), the free energy is given by
\begin{equation}
\begin{split}
F=&-\frac{1}{\beta} \sum_{\nu} \ln \left( 1+ e^{- \beta E_{\nu}^{\uparrow}}\right) + \sum_{i} \frac{|\Delta_{i}|^{2}}{V} -\sum_{i} (\mu + h)  \label{6}
\end{split}
\end{equation}
Here, we just compare the summation of the first two terms, because the third term is a constant for all solutions of different phases.
In the following we determine the phase boundary between FF state, LO state, and normal state.

\subsubsection {First order transition between the FF and LO phases}

\begin{figure}[ht]
\begin{center}
\includegraphics[width=8cm, clip]{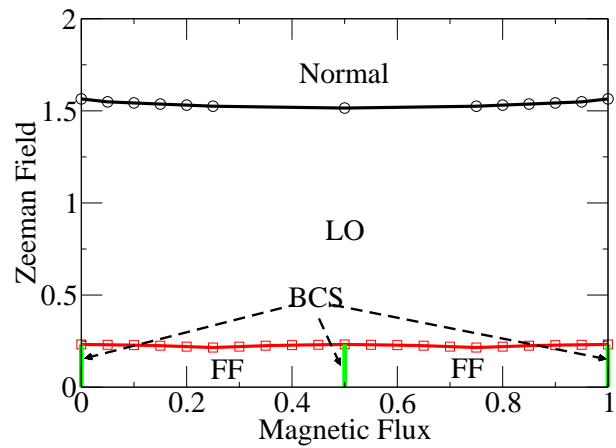}
\end{center}
\caption{(Color online) Phase diagram of the superconducting ring in the $h$-$\Phi$ plane. Here the ring size is $N=50$, and the temperature is $T=0.005 $ ($\beta=200$). Notice that the boundary line between the FF and LO phases has the periodicity in $\Phi$ with a period of $\Phi_0/2$ while that between the LO and normal state phases has the periodicity in $\Phi$ with a period of $\Phi_0$.
Specifically, the LO state exists in the range $[0.23,1.56]$, $[0.21,1.52]$, and $[0.23,1.51]$
 for $\Phi=0$,  $\Phi_0/4$, and $\Phi_0/2$, respectively.}
\label{fig2}
\end{figure}

When determining the pair potential self-consistently by iteration, we find that in certain range of the strength of the in-plane Zeeman field $h$, different initial configurations of the pair potential lead to different stable solutions. In another word, there are more than one stable solutions through iteration. For example, when we fix $\Phi= \Phi_{0}/4$, and vary the magnetic field in the range $0.08<h<0.29$, stable solutions of both the FF type and the LO type can be arrived at through iteration. The free energies of these two types of stable solutions are listed in Table I. It can be seen that the LO state becomes energetically favorable when the magnetic field is equal to or greater than $h_{c1}=0.21$. In addition, the free energy at $h_{c1}$ is continuous, but its first order derivative is not continuous. Hence, we conclude that for a fixed magnetic flux $\Phi= \Phi_{0}/4$, there is a first-order phase transition between the FF and LO states at $h_{c1}$. Similarly, we fix magnetic flux $\Phi$ at different values and we can find the threshold value of $h$ at which the system changes from the FF state to the LO state or vise versa. Thus for a fixed temperature $\beta=200$ and fixed system size $N=50$, the phase transition line between FF and LO state is determined by comparing the free energy of the FF phase and the LO phase, and we plot it in Fig.~\ref{fig2}. To ensure that the phase boundary given by $N=50$ is close to that of the thermodynamic limit, we change the system size to $N=200$, and we find the phase boundary does not change. For $N=50$ and $N=200$, the magnitude of the OP $\Delta_{i}$ in BCS phase is the same. Hence the result based on $N=50$ can be regarded as in thermodynamic limit. It can be seen that the first-order transition line is not parallel to the $\Phi$ axis, so we can turn the flux to make the system change from the LO phase to the FF phase or vise versa. We call this phase transition AB effect induced phase transition. We can also see that the phase boundary between the LO and FF states is symmetric around $\Phi= \Phi_{0}/4$, and the period of FF phase is $\Phi_{0}/2$.
\begin{table}[tbp]
\caption{Free energies (up to a constant $-\sum_{i=1}^{N} (\mu + h)$) for stable solutions of FF state and LO state. Here the ring size is $N=50$, the magnetic flux $\Phi=\Phi_{0}/4$, and the temperature $\beta=200$. It can be seen that there is a first-order phase transition from the FF state to LO state when the in-plane magnetic field is tune across $h=0.21$}
\begin{tabular}{c|c|c|c|c|c|c|c}
\hline\hline
& $h=0.20$ & $h=0.21$ & $h=0.22$ & $h=0.23$ & $h=0.24$ & $h=0.25$ \\ \hline\hline
\parbox{0.5cm} {FF} &
\parbox{1.2cm} {-56.1487}&
\parbox{1.2 cm} {-55.6487} &
\parbox{1.2 cm}{-55.1487}  & \parbox{1.2cm} {-54.6487}& \parbox{1.2cm} {-54.1487}&\parbox{1.2cm} {-53.6487}  \\ \hline
\parbox{0.5cm} {LO} &
\parbox{1.2cm} {-56.1019} &
\parbox{1.2 cm} {-55.6319} &
\parbox{1.2 cm} {-55.1619}  &\parbox{1.2cm} {-54.6919} & \parbox{1.2cm} {-54.2219}&\parbox{1.2cm} {-53.7519}  \\ \hline\hline
\end{tabular}
\end{table}

\subsubsection {Second-order transition between the LO and normal state phases}

If we continue to increase the in-plane magnetic field above  the value $h=0.29$ for $\Phi$ fixed at $\Phi_0/4$,
all initial configurations of the pair potential will lead to the LO state, or only the LO state becomes stable. Meanwhile the amplitude of the pair potential decreases and the period of the modulation of the pair potential is shortened continuously. Further increase of the Zeeman field leads to the reduction of the pair potential until it vanishes gradually. When the magnetic field reaches $h_{c2}=1.52$, the amplitude of the pairing potential vanishes, or the LO state is completely depressed by the in-plane magnetic field, and the system changes from the LO state to the normal state. If we do the iteration from zero pair potential, we will find that when $h > h_{c2}$ the stable solution for the pair potential is zero (normal state). When $h\leqslant h_{c2}$, the stable solution is an LO state. There is no coexistence area of the LO and the normal states in the $h$ axis. Hence we conclude that the phase transition at $h_{c2}$ for a fixed $\Phi$ is a second-order phase transitoin. Our result is consistent with previous
studies.~\cite{YMatsuda07,HShimahara94,KYang98}

\subsection {Phase boundary in $\emph{h}$-$\emph{T}$ plane}

\begin{figure}[ht]
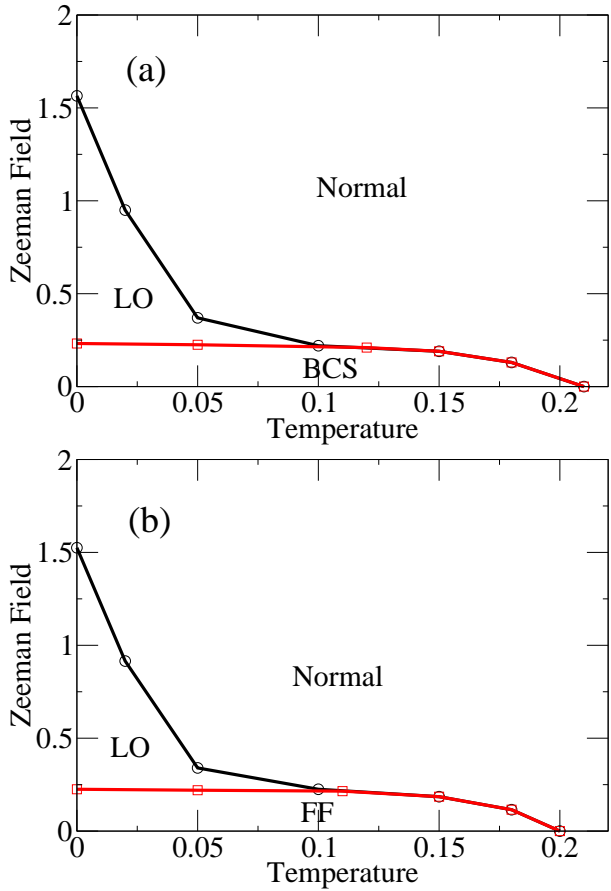

\begin{center}
\includegraphics[width=8cm, clip]{fig3a.eps}
\includegraphics[width=8cm, clip]{fig3b.eps}
\end{center}
\caption{(Color online) Phase diagram of the superconducting ring in the $h$-$T$ plane. Here the ring size is $N=50$, and the magnetic flux are $\Phi=0$ (a), and $\Phi=\Phi_{0}/4$ (b), respectively. For $\Phi=0$, the LO state emerges at low temperature $T<0.12$ and relatively high magnetic field $0.23<h<1.56$. For $\Phi=\Phi_0/4$, the LO state emerges at low temperature $T<0.11$ and relatively high magnetic field $0.21<h<1.52$.
The LO state to normal state transition (Black with open circles) is of second order, while the BCS (or FF) to the normal state transition (Red with open squares) is of first order.
The zero-field transition temperature is around $T_{c}=0.21$ and $0.20$ for $\Phi=0$ and $\Phi_0/4$. }
\label{fig3}
\end{figure}

In the preceding subsection, we study the phase transitions when we vary the magnetic flux $\Phi$ or the in-plane magnetic field $h$. The temperature is fixed at a very low value. Hence these phase transitions can be regarded as quantum phase transitions. In this subsection, we will study the phase transitions induced by thermal fluctuations, and determine their phase boundaries. We will fix the magnetic flux $\Phi$ and vary the temperature $\beta$ or the in-plane magnetic field $h$. First we consider the case in the absence of the magnetic flux $\Phi=0$. We fix the temperature at $\beta$=20 ($T$=0.05), $\beta$=10 ($T$=0.10), $\beta$=6.67 ($T$=0.15), $\beta$=5 ($T$=0.20), and $\beta$=4 ($T$=0.25) respectively, and do the iteration separately. The phase boundary between the BCS and LO states is determined in a similar way to that in Sec.~\ref{sec:result}.A. It can be seen that when we fix the magnetic flux to be zero, and tune the in-plane magnetic field or the temperature, the system will change between the BCS, LO, and normal states. As can be seen from  Fig.~\ref{fig3}(a). the LO phase emerges below the critical temperature $T\approx 0.12$.
We note  that the BCS to the LO state is first order, and the LO to normal is second order. When the magnetic flux is nonzero, the BCS state will be replaced by FF state with the phase diagram,
as shown in Fig.~\ref {fig3}(b), very similar to the zero-flux case.

\subsection {Mesoscopic effect}

\begin{figure}[ht]
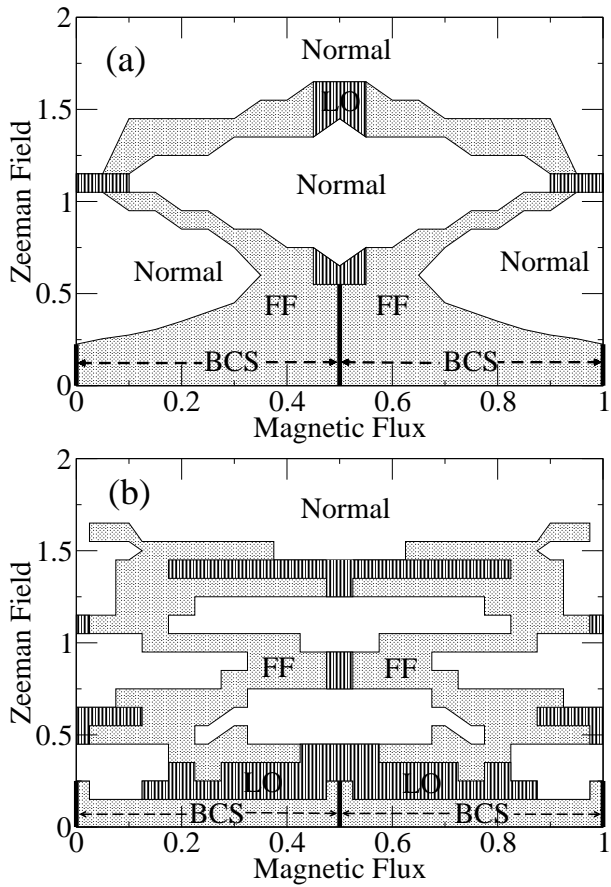

\begin{center}
\includegraphics[width=8cm, clip]{fig4a.eps}
\includegraphics[width=8cm, clip]{fig4b.eps}
\end{center}
\caption{Phase diagram in the $h$-$\Phi$ plane. All the parameters are the same as
that in Fig.~\ref {fig2} except that the ring size is $N=20$ (a) and $N=10$ (b).
The empty area represents the normal state. The gray area represents the FF state, and the area covered by the thin black lines represents the LO state. The thick black lines
represent the BCS state. It can be seen that with the decrease of the ring size, the FF phase expand a lot and the LO phase shrink dramatically. The period of FF phase also changes from $\Phi_{0}/2$ to $\Phi_{0}$}
\label{fig4}
\end{figure}

Another interesting question is the mesoscopic effect. In this subsection we will study the mesoscopic effect by fixing the temperature and decreasing the ring size. As mentioned in the above discussion, the ring size $N=50$ already gives the same phase boundary as that of $N\rightarrow \infty$. A simple check is that when we increase the ring size to $N=100$, and $N=200$, we find that the phase boundaries do not change in comparison with that for $N=50$. This means that for the current model, $N=50$ can be treated as in the thermodynamic limit. However, if we decrease the ring size, for example to $N=20$, the mesoscopic effect will occur. First, in the $h$-$\Phi$ plane, the LO phase will shrink dramatically and the FF phase will expand (see Fig.~\ref{fig4}(a)). This is because (1) the influence of the magnetic flux on the system will increase and the influence of the in-plane Zeeman field will decrease relatively, and (2) having a finite size restricts the periodicity of the LO order parameter. At a given Zeeman field, if the period is not commensurate with the corresponding ring, solutions of the LO state will have to be modified to be commensurate with system size, which results in some energy cost. Therefore,
 the LO state will shrink. Second, the periodicity of the magnetic flux changes from $\Phi_{0}/2$ to $\Phi_{0}$. This is because the system size is so small that the Cooper pair can no longer be treated as a whole, and can only be treated as two separate electrons. We show in Fig.~\ref{fig4}(b) the phase boundary for $N=10$ and the re-entrant behavior of various phases can be seen in the phase diagram.

\section {Discussion and conclusion}
\label{sec:discussion}

Based upon a tight-binding model, we study a one-dimensional $s$-wave superconducting ring subject to an in-plane Zeeman field and a magnetic flux by solving  the BdG equation in real space. In the presence of a magnetic flux,  a crossover from the BCS state to the FF state is obtained when the in-plane magnetic field is not very strong. If we increase the strength of the in-plane magnetic field, the LO state becomes favorable, and a FF to LO phase transition occurs. With the further increase of the in-plane magnetic field strength, the magnitude of the pair potential of the LO state is suppressed, and disappears finally with the system entering the normal state. In the absence of the magnetic flux, there is no FF phase, and the Zeeman field induces the
transitions between the BCS, LO, and normal states, which has been studied
extensively.~\cite{HShimahara94,KYang98,YTanuma98,JXZhu00,DFAgterberg01,
QWang06a,QWang06b,QCui08,YYanase07,TDatta09}
Our results agree well with the previous studies in a two-dimensional system that the energetically favorable state for $s$-wave superconductor is a one-dimensional stripe-like LO state. This  suggests the first-order transition between the BCS and LO states while a second-order transition between the LO to normal states. Our study goes beyond that and indicates a stable FF state due to the magnetic flux. The mesoscopic effects are also studied. When the system size decreases, two mesoscopic effects arise: (1) the LO phase in the $h$-$\Phi$ plane shrinks, and the FF state expands due to the enhancement of the Aharanov-Bohm effect; (2) the periodicity of the external magnetic flux will change from $\Phi_0/2$ to $\Phi_0$.

The following remarks are in order: (1) Though we study a one-dimensional model, the system should not be regarded as a mathematically one-dimensional. The current study can be easily extended to the two-dimensional and other geometry, such as a torus configuration threaded by a magnetic flux. It can be expected that a similar phase transition between LO state and FF state will occur.
(2) For the one-dimensional case,  the LO state exists in a broader range of parameters space ($h-T$ space, see Fig.~\ref{fig2}) than that of two-dimensional and
three-dimensional cases,~\cite{YMatsuda07,YSuzumura83,KMachida84} which makes it easier to access experimentally.
(3) In Ref.~\onlinecite{KHenderson09}, it is reported that an trap potential with arbitrary configuration can be achieved. Hence, we expect that the result presented in this paper should be able to be observed experimentally in cold Fermions under current experiment technique.  (4) In the thermodynamic limit, $N\rightarrow \infty$, the FF state reproduces the BCS state, because the phase gradient of the order parameter is vanishingly small. This result agrees with our intuition that when the ring size becomes infinity, the influence of the magnetic flux can be neglected. (5) The effect of the impurity is not included in the current study, and will be given in our future studies.

\begin{acknowledgments}
One of us (H.T.Q.) thanks Rishi Sharma for stimulating discussions. This work was supported by
U.S. DOE at LANL under Contract No. DE-AC52-06NA25396, the U.S. DOE Office of Science, and the LANL LDRD Program.
\end{acknowledgments}

\end{document}